\documentclass[aps,prc,twocolumn,showpacs,graphicx]{revtex4}
\usepackage[dvips]{graphicx}

\def\beq{\begin{equation}}
\def\eeq{\end{equation}}
\def\bea{\begin{eqnarray}}
\def\eea{\end{eqnarray}}

\begin{document}

\noindent
\title{Spin-orbit correlation energy in neutron matter}


\author{M. Baldo}
\affiliation{Istituto Nazionale di Fisica Nucleare, Sezione di Catania,
Via S. Sofia 64, I-95123 Catania, Italy} 
\author{C. Maieron}
\affiliation{Istituto Nazionale di Fisica Nucleare, Sezione di Catania,
Via S. Sofia 64, I-95123 Catania, Italy}

\date{\today}

\bigskip

\begin{abstract}
We study the relevance of the energy correlation produced by
the two-body spin-orbit coupling present in realistic nucleon-nucleon
interaction potentials. To this purpose, the neutron matter
Equation of State (EoS) is calculated with the realistic two-body 
Argonne $v_8'$ potential. The shift occurring in the EoS when
spin-orbit terms are removed is taken as an estimate of the
spin-orbit correlation energy. Results obtained within the
Bethe-Brueckner-Goldstone expansion, extended up to three hole-line
diagrams, are compared with other many-body calculations
recently presented in the literature. In particular, excellent
agreement is found with the Green's function Monte-Carlo method.     
This agreement indicates the present theoretical accuracy in the calculation
of the neutron matter EoS.
 
\end{abstract}

\bigskip
\bigskip
\bigskip


\maketitle

\section{Introduction}

The  properties of nuclear matter  at high density play a crucial role
in the modeling of neutron stars (NS's) interior \cite{shapiro}. 
The  observed NS  masses are in the range of  
$\approx (1-2)  M_\odot$  (where  $M_\odot$  is  the  mass of the sun, 
$M_\odot =
1.99\times 10^{33}$g), and the radii are of the order of 10 km. 
 The matter inside NS's, below the outer crust,  
possesses densities ranging from a fraction to a few times 
the normal nuclear matter density  $\rho_0$
$(\approx  0.17\;{\rm  fm}^{-3})$.
The equation of state (EoS) at such densities is one of the
main ingredients to determine the structure parameters of NS's.
Due to beta-stability conditions, NS matter is much closer to 
neutron matter than to symmetric nuclear matter \cite{shapiro}. 
Since no phenomenological
data can be used to constrain the neutron matter EoS, one has to rely on 
microscopic many-body calculations based on realistic nucleon-nucleon (NN)
interactions. Predictions of the neutron matter EoS based on 
purely phenomenological Skyrme forces can dramatically differ among each 
other, even at relatively low density (for a recent
compilation see {\it e.g.} \cite{ABrown}). For these reasons,
an accurate determination of the neutron matter
EoS in the density range typical of NS's, based on many-body
theory and realistic NN forces, appears of great relevance. 
Unfortunately, despite the enormous
progress in the many-body theory of nuclear matter in general and
neutron matter in particular, discrepancies among different
calculations are still persisting. Among the variety of methods that have been
developed in the many-body theory of nucleon systems, one can
mention the variational method \cite{Akmal}, 
in his various degrees of sophistication
(including Monte-Carlo procedures), the Green' s function Monte-Carlo 
method (GFMC) \cite{Pieper}, which represents a numerical algorithm
converging in principle to the ``exact" solution, 
and the Bethe-Brueckner-Goldstone
(BBG) expansion \cite{book}. It has been pointed out \cite{Fantoni} that
the spin-orbit interaction and correlations are particularly relevant
in neutron and nuclear  matter and require accurate many-body and 
numerical treatments.
Indeed, a large fraction of the observed discrepancies 
seems to reside in the proper treatment of the spin-orbit interaction terms.
\par
In this paper we focus on the many-body effects due to the spin-orbit 
terms of the NN interaction within the BBG expansion. To this purpose,
and for the sake of comparison with the results obtained with other methods,
we perform calculations for neutron matter with the $v_8^\prime$ and 
$v_6^\prime$ NN realistic two-body interactions, including and, respectively,
not including the spin-orbit terms. These interactions are
simplified versions of the $Av_{18}$ potential, but they can be still 
considered realistic enough to provide meaningful results. 
In particular, they both contain tensor interaction operators.   
This paper is organized as follows. 
In Sec.~II we briefly introduce the BBG expansion method and
discuss the corresponding  neutron matter EoS in the relevant density range.
In Sec.~III we make a detailed comparison with the results of
other many-body methods, in particular the GFMC and the Auxiliary
Field diffusion Monte-Carlo method (AFDMC). Sec.~IV is devoted
to the conclusions.

\section{Neutron matter EoS and the BBG expansion}

The main difficulty in the many-body theory of nuclear matter 
is the treatment of the 
strong repulsive core, which dominates the short range behavior of 
the NN interaction.
Simple perturbation theory cannot of course be applied,
and one way of overcoming this difficulty is to introduce the two-body
scattering G-matrix, which has a much smoother behavior even for
a large repulsive core. It is possible to rearrange the perturbation
expansion in terms of the reaction G-matrix, in place of the original
bare NN interaction, and this procedure is systematically exploited
in the BBG expansion \cite{book}.
\par
The expansion of the ground state energy at a given
density, i.e. the EoS at zero 
temperature, can be ordered according to the number of independent
hole-lines appearing in the diagrams representing the different
terms of the expansion. This grouping of diagrams generates the
so-called hole-line expansion \cite{Day}.
The diagrams with a given
number $n$ of hole-lines are expected to describe the main contribution
to the $n$-particle correlations in the system. At the two
hole-line level one gets the Brueckner-Hartree-Fock (BHF) approximation.
The BHF approximation includes the self-consistent
procedure of determining the single particle auxiliary potential,
which is an essential ingredient of the method.
Once the auxiliary self-consistent potential is introduced,
the expansion is implemented by introducing the set of diagrams
which include ``potential insertions" \cite{book}. 
To be specific, the introduction 
of the auxiliary potential can be formally performed by splitting the
hamiltonian in a modified way from the usual one
\beq
H = T + V = T + U + (V - U) \equiv H_0' + V' \;,
\label{eq:split}
\eeq 
\noindent
where $T$ is the kinetic energy and $V$ the nucleon-nucleon 
interaction. Then one considers $V' = V - U$ as the new interaction potential
and $H_0'$ as the new single particle hamiltonian.
The modified single particle energy $e(k)$ is given by
\beq
e(k) = {\hbar^2 k^2\over 2m} + U(k)
\label{eq:spe}
\eeq
\noindent
while $U$ must be chosen in such a way that the new interaction $V'$
is, in some sense, ``reduced" with respect to the original one $V$, so that
the expansion in $V'$ should be faster converging. The introduction of the
auxiliary potential turns out to be essential, otherwise the 
hole-expansion would be badly diverging. The total energy $E$ can then
be written as 
\beq
 E = \sum_k e(k) + B
\label{eq:ene}
\eeq
\noindent
where $B$ is the interaction energy due to $V'$. 
The first potential insertion diagram
cancels out the potential part of the single particle energy of 
Eq. (\ref{eq:spe}),
in the expression for the total energy $E$. This is actually true
for any definition of the auxiliary potential $U$. At the two 
hole-line level of approximation, one therefore gets
\beq
   E = \sum_{k < k_F}  {\hbar^2 k^2\over 2m} + {\tilde B}
\label{eq:enep} 
\eeq

The result that only the unperturbed kinetic energy appears in the
expression for $E$ and that all correlations are included in the
potential energy part $\tilde{B}$  holds true to all orders and 
it is a peculiarity
of the BBG expansion. Of course, the modification of the momentum 
distribution, and therefore of the kinetic energy, is included in the
interaction energy part, but it is treated on the same footing as the
other correlation effects. This presents a noticeable
advantage: in fact, the modification of the kinetic energy itself
is quite large and, of course, positive and it should be therefore 
compensated by an extremely accurate calculation of the 
(negative) correlation energy, if the two quantities 
are calculated independently.\par
\begin{figure}
\vspace{-3. cm}
  \includegraphics[width=0.7\textwidth]{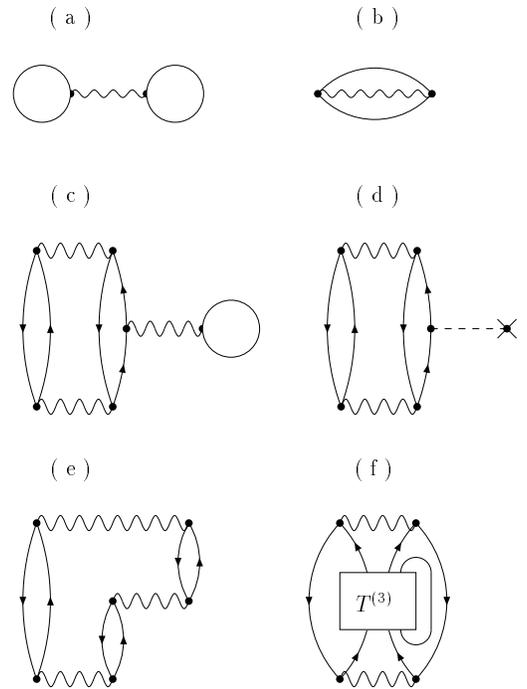} 
\vspace{-8. cm}
\caption{Schematic representation of two hole- and three hole-line
diagrams. Both direct and
 exchange diagrams are included. The wavy line indicates 
 a Brueckner G-matrix, the dotted line an U-insertion. For other details,
see text.}
\label{fig:3h}
\end{figure}

Up to three-hole lines, the diagrams for ${\tilde B}$ can be schematically 
represented
as in Fig.~\ref{fig:3h}. Diagrams (a) and (b)
in the first line represent the usual BHF approximation,
while the remaining lines include the
three hole-line diagrams.
\noindent
The box in the third line represents
the three-body scattering matrix $T^{(3)}$, which can be introduced following 
a procedure similar to what is done for the Brueckner G-matrix and 
satisfies the Bethe-Fadeev
integral equations \cite{raja, book}.
Diagram (f) generates, to lowest order
in the G-matrix, diagrams (c) and (e), which are usually calculated
separately, since they require an accurate numerical procedure  
(and they cancel each other to a large extent). Diagram (d) is
a potential insertion diagram, the only one at the three hole-line level,
and it is non-zero only if the single particle potential is non-zero
at momenta larger then the Fermi momentum $k_F$ 
({\it e.g.} in the so-called ``continuous choice" \cite{Mahaux,book}).
\par
In a previous paper \cite{Baldo2000} 
we have shown that the BBG expansion for 
neutron matter displays a relatively rapid rate of convergence, and 
that calculating the total energy up to the three hole-line diagrams is enough
to get an accurate EoS, even for densities a few times larger than 
the saturation density. These calculations were done for the Argonne
$v_{14}$ and $v_{18}$ potentials, which contain a large set of interaction
operators, including the spin-orbit ones. In order to simplify
the analysis and the comparison with other methods, we have performed the
same type of calculations for the $v_{8}^{\prime}$ NN potential
\cite{v8prime},
which is a simplified version of the $v_{18}$ interaction, but still
realistic enough to keep the calculations meaningful. 
To better focus on the effects of the spin-orbit interaction,
we have also considered the $v_6^\prime$ potential, which is obtained
from $v_{8}^{\prime}$ by dropping the spin-orbit terms. 
The difference of the EoS 
obtained with these two interactions can, therefore, be taken as an 
estimate of the contribution and relevance of the spin-orbit 
NN interaction in neutron matter.\par
\begin{figure}[t]
  \includegraphics[width=0.5\textwidth]{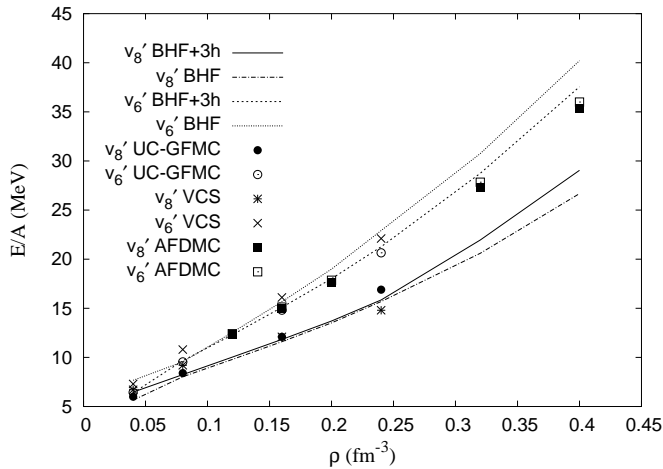} 
   \caption{EoS of neutron matter. The dotted and dash-dotted lines
    correspond to BBG calculations at the BHF level for the $v_6'$ and 
    $v_8'$ NN interactions, respectively, while the short dashed ($v_6'$) 
    and solid ($v_8'$) lines include three hole-line contributions.
     Empty ($v_6'$) and full ($v_8'$) circles are the unconstrained
     GFMC results of Ref.~\cite{Carlson} and crosses ($v_6'$) and 
     stars($v_8'$) are the VCS results of Ref.~\cite{Carlson}. 
     Finally empty ($v_6'$) and full ($v_8'$) squares represent the 
     AFDMC results of Ref.~\cite{Fantoni}. See text for a discussion
     of finite size corrections applied to the variational and 
     Monte Carlo results.
}
\label{fig:eos}
\end{figure}
The EoS of neutron matter obtained within the BBG
expansion for the two considered NN interactions is shown 
in Fig. \ref{fig:eos}. 
All calculations have been performed in the continuous 
choice for the single particle potential $U(k)$. As shown in 
Ref.~\cite{Baldo2000},
the results are independent, to a high degree of accuracy, of the
choice of $U(k)$.
At the BHF level the difference between
the EoS for the $v_{8}'$ NN interaction (dash-dotted line) and
the EoS for the $v_{6}'$ NN interaction (dotted line) looks
sizable, even at relatively low density. This discrepancy 
increases with density, reaching about
14 MeV at $\rho = 0.4$ fm$^{-3}$. 
When three-body correlations are included (dashed and full lines),
the gap between the two EoS is reduced,
the strength of this reduction being 
about 40\% at the highest considered density. 
\par
Notice that the contribution
of three hole-lines diagrams is positive for $v_{8}'$ and negative
for $v_{6}'$, therefore this difference turns out to be quite sensitive 
to three-body correlations (as defined within the BBG expansion).  
\par
It is important to stress that the contribution of three-body correlations
is only a few percent with respect to the two-body one (BHF),
even at the highest density. Indeed, in the considered density region 
the interaction 
energy ${\tilde B}$ of Eq. (\ref{eq:enep}) is 
negative and large and compensates
a large fraction of the positive kinetic energy contribution.  
This substantial cancellation between kinetic energy and interaction energy
has the effect of amplifying the (small) differences of the
correlation energy obtained in the different EoS 
calculations for a given NN interaction, as discussed in the next Section.

\section{Results and Discussion}
BBG results for the EoS of neutron matter using the $v_8^\prime$ 
and $v_6^\prime$ interactions are compared in Fig.~\ref{fig:eos} 
and, respectively, in Tables \ref{tab:v8} and \ref{tab:v6}
with other calculations based on different many-body methods.
\par
Green' s function Monte-Carlo results 
were obtained in Ref.~\cite{Carlson} within an unconstrained path (UC)
approach, by considering 14 neutrons inside a periodic box and setting the
interaction potential discontinuously to zero at distances larger than
one-half the box size. The GFMC should give, in principle, the exact 
ground state energy of the system. For comparison to infinite neutron 
matter  a ``box correction"  must then be applied.
For the $v_8'$ interaction the latter was estimated in
Ref.~\cite{Carlson} 
using  variational chain summation 
(VCS) techniques and turned out to be mostly due to the truncation of the
potential.  
The resulting box corrected UC-GFMC EoS is listed in the fourth 
column of Table~\ref{tab:v8} and indicated in Fig.\ref{fig:eos} by 
full circles. Up to the highest density considered in \cite{Carlson}, 
the agreement
with the BBG results looks remarkable, given the
uncertainty contained in the box correction procedure.
\par 
The authors of Ref.~\cite{Carlson} also calculated 
the neutron matter EoS 
within the variational chain summation approach, both 
for 14 neutrons in a periodic box and directly for an uniform 
gas of neutrons.
Their results for the infinite system with the $v_8'$ NN interaction are 
listed in the last column of 
Table~\ref{tab:v8} and plotted as stars in Fig.~\ref{fig:eos}.
They show fairly good agreement with the other methods,  
except, maybe, for the last two points at higher density, which seem to display
a slightly different slope with respect to UC-GFMC and BBG
\footnote{The different density dependence of the UC-GFMC and VCS results
when the spin-orbit interaction is included in the NN potential
was observed, already in the periodic box results, in Ref.\cite{Carlson},
where the possible origin of this behavior was supposed to be due 
either to an overestimation of
spin-orbit contributions in the VCS method or to the employment
of a too short imaginary time in the UC-GFMC calculation. The slightly
better agreement between UC-GFMC and the trend of the BBG curves shown in 
Fig.~\ref{fig:eos} seems to favor the first hypothesis.}.
\par
Switching the spin-orbit interaction terms off and considering the 
$v_6^\prime$ NN potential, we can again compare the BBG results with UC-GFMC 
and variational calculations, which are also available \cite{Carlson}. 
Unfortunately in this case the box correction to the GMFC has not been 
provided; however it looks reasonable to use the same 
corrections as in the 
$v_8^\prime$ case. Once these corrections
are applied to the UC-GMFC of Ref.~\cite{Carlson} 
for the $v_6^\prime$ NN interaction, we obtain the
neutron matter EoS listed in 
Table~\ref{tab:v6} and indicated by the 
open circles in Fig.~\ref{fig:eos}. Again, fairly 
good agreement with the BBG calculations is observed up to the
highest density, $\rho = 0.24$ fm$^{-3}$,  considered in ~\cite{Carlson}.
The overall trend seems to indicate that this agreement
continues also at higher $\rho$ values. 
\par
It is interesting to notice that the inclusion of three-body correlations 
in the BBG expansion plays a relevant role in improving 
the agreement with the UC-GMFC 
results, which is, in any case, already satisfactory at the 
BHF level. This happens
both for the $v_8'$ potential, for which, as already mentioned, the
contribution of three hole-line diagrams is positive, and for the $v_6'$
case, where, instead, the correction is negative.
\par
The variational VCS results, plotted as crosses in Fig.~\ref{fig:eos},
have been obtained also from  Ref.~\cite{Carlson}, again correcting
periodic box $v_6'$ results with the box correction values given
for the $v_8^\prime$ interaction. Also in this case the agreement
can be considered satisfactory, and again the trend with density
seems to be slightly different with respect to UC-GFMC and BBG. 
\par
Another Monte-Carlo scheme has been recently developed \cite{Fantoni} 
for neutron matter, the Auxiliary Field Diffusion Monte-Carlo (AFDMC)
method. 
\par
The recent results of Ref.~\cite{Fantoni} for 
$v_6^\prime$ 
are reported in Fig.~\ref{fig:eos} as open squares.
These calculations were performed for 14 neutrons in a cubic box,
but using a continuous potential instead of the truncation
of Ref.~\cite{Carlson} and they therefore  
should automatically incorporate
the largest part of finite size effects. 
As shown in Table~\ref{tab:v6}, these results are close to
the calculated UC-GFMC EoS, as well as to the BBG EoS 
in the whole considered density range.
\par
Unfortunately, the same 
method applied to neutron matter with the $v_8^\prime$ NN 
interaction gives an EoS (full squares in Fig.~\ref{fig:eos}
and third column in Table~\ref{tab:v8}) which differs 
from all other calculations. These findings indicate the difficulty of
an accurate calculation of the spin-orbit contribution
to neutron matter binding.
The AFDMC method has been 
improved in Ref.~\cite{Fantoni2} to properly deal with the spin-orbit
interaction and correlation, by a suitable modification  (backflow)
of the trial wave function. With this modification the splitting between
the two EoS, for the $v_6^\prime$ and $v_8^\prime$ NN 
potentials, increases, but 
it is still too small with respect to the UC-GFMC and BBG results.  
\begin{table}[t]
    \caption{Neutron matter energy per particle (in MeV), for the
      Argonne $v_8'$ NN interaction. BBG energies are calculated
      up to the three-hole line level. AFDMC results are taken from 
      Ref.~\cite{Fantoni} and are calculated for 14 neutrons in a 
      periodic box. UC-GFMC and VCS results are taken 
      from Ref.~\cite{Carlson}. 
      VCS results are calculated for a Uniform Gas, while UC-GFMC results 
      include box correction terms.}
\begin{ruledtabular}
\begin{tabular}{ccccc}
$\rho$ (fm$^{-3}$) & BBG    & AFDMC  & UC-GFMC  & VCS   \\
\hline
 0.04 &              6.469  &   --   &  6.0  &  6.7  \\
 0.08 &              8.250  &   --   &  8.4  &  9.2  \\
 0.12 &             10.031  &  12.32 &   --  &   --  \\
 0.16 &             11.826  &  14.98 & 12.1  & 12.1  \\
 0.20 &             13.705  &  17.65 &  --   &  --   \\
 0.24 &             15.846  &   --   & 16.9  & 14.8  \\
 0.32 &             21.953  &  27.3  &  --   &  --   \\
 0.40 &             29.044  &  35.3  &  --   &  --   \\
\end{tabular}
\end{ruledtabular}
\label{tab:v8} 
\end{table}
\begin{table}[t]
\caption{Same as Table~\protect\ref{tab:v8}, but for the
Argonne $v_6'$ NN interaction. UC-GFMC and VCS energies are obtained 
from the periodic 
box results of Ref.~\cite{Carlson}, by applying the same box correction
(reported in the last column) as for the $v_8^\prime$ case}
\begin{ruledtabular}
\begin{tabular}{cccccc}
$\rho$ (fm$^{-3}$) & BBG    & AFDMC & UC-GFMC & VCS  & (box corr.) \\
\hline
 0.04              &  6.389 & --    & 6.45    & 7.3 & (-0.3)      \\
 0.08              & 9.668  & --    & 9.54    & 10.8 & (-1.1)      \\
 0.12              & 12.292 & 12.41 & --      & --   & --          \\
 0.16              & 15.092 & 15.12 & 14.81   & 16.1 & (-5.1)      \\
 0.20              & 18.011 & 17.86 & --      & --   & --          \\
 0.24              & 21.262 & --    & 20.65   & 22.1 & (-11.5)     \\ 
 0.32              & 28.743 & 27.84 & --      & --   & --          \\
 0.40              & 37.552 &  36.0 & --      & --   & --          \\
\end{tabular}
\end{ruledtabular}
\label{tab:v6} 
\end{table}

\section{Conclusions}   
We have presented calculations of neutron matter EoS for two different
NN interactions, the Argonne $v_6^\prime$ and $v_8^\prime$. The comparison of
the results obtained with the two interactions is expected to give an
estimate of the correlation coming from spin-orbit interaction terms.
We have found close agreement between the EoS calculated 
within the BBG expansion, 
extended up to three hole-line contributions, and the UC-GFMC calculations
of Ref.~\cite{Carlson}, for density up to 0.24 fm$^{-3}$. The discrepancy 
between the correlation energy in the two schemes does not
exceed 2\%.  Such an agreement suggests that the many-body problem
for neutron matter with two-body NN interactions is well under control,
at least for the considered density range. The splitting between the
EoS calculated with the $v_6^\prime$ and $v_8^\prime$ 
potentials indicates that the
spin-orbit correlation energy in neutron matter can be as large as 4-5 Mev/A 
and increases with density. The AFDMC methods seems to have some
problems when dealing with the spin-orbit correlation. \par
In all considered calculations only two-body NN forces have been considered.
It is well known that three-body forces are needed in nuclear matter.
It appears then relevant to perform a similar 
study including three-body forces.
The latter are not so well known, 
and the extrapolation from finite nuclei, where 
three-body forces are fitted, to nuclear matter 
seems not so obvious \cite{Fantoni}. 
In any case, the comparison of the results obtained
with different schemes, when three-body forces are included,
could be a stringent test for the accuracy of the calculations,
in particular for the spin-orbit contribution to correlation energy.
This is left to future work, but the agreement obtained up to now
between BBG and GFMC appears satisfactory and promising.


\end{document}